\documentclass[twocolumn,showpacs,preprintnumbers,amsmath,amssymb,superscriptaddress,pra]{revtex4}
\pdfoutput=1
\usepackage{graphicx}% Include figure files
\usepackage{dcolumn}% Align table columns on decimal point
\usepackage{bm}% bold mathCava
\usepackage{wrapfig}% wrapping text around a figure
\usepackage{amssymb}
\usepackage{amsmath}
\usepackage{color}

\usepackage{pdfsync}

\def\be{\begin{equation}}
\def\ee{\end{equation}}
\def\bea{\begin{eqnarray}}
\def\eea{\end{eqnarray}}

\newcommand{\ie}{{\it{i.e.~}}}
\newcommand{\etal}{{\it{et al.}}}

\newcommand{\ket}[1]{| #1 \rangle}
\newcommand{\bra}[1]{\langle #1 |}

\begin{document}

%preprint{APS/123-QED}

\title{Noisy evolution of graph-state entanglement}
\author{L. Aolita}
\affiliation{ICFO-Institut de Ciencies Fotoniques, Mediterranean
Technology Park, 08860 Castelldefels (Barcelona), Spain}
\author{D. Cavalcanti}
\affiliation{Centre for Quantum Technologies, University of Singapore, Singapore}
\author{R. Chaves}
\affiliation{Instituto de F\'\i sica, Universidade Federal do Rio
de Janeiro. Caixa Postal 68528, 21941-972 Rio de Janeiro, RJ,
Brasil}
\author{C. Dhara}
\affiliation{ICFO-Institut de Ciencies Fotoniques, Mediterranean
Technology Park, 08860 Castelldefels (Barcelona), Spain}
\author{L. Davidovich}
\affiliation{Instituto de F\'\i sica, Universidade Federal do Rio
de Janeiro. Caixa Postal 68528, 21941-972 Rio de Janeiro, RJ,
Brasil}
\author{A. Ac\'in}
\email{antonio.acin@icfo.es}
\affiliation{ICFO-Institut de
Ciencies Fotoniques, Mediterranean Technology Park, 08860
Castelldefels (Barcelona), Spain}
\affiliation{ICREA-Instituci\'o
Catalana de Recerca i Estudis Avan\c cats, Lluis Companys 23,
08010 Barcelona, Spain}

\begin{abstract}
A general method for the study of the entanglement evolution of graph states under the action of Pauli maps
was recently proposed in [Cavalcanti \etal, Phys. Rev. Lett. {\bf 103},
030502 (2009)]. This method is based on lower and upper bounds to the entanglement of the entire state as a function only of the state
of a (typically) considerably-smaller subsystem undergoing an effective
noise process related to the original map. This provides a huge decrease in the size of the matrices involved in the calculation of entanglement in these systems. In the present paper we elaborate on this method in
detail and generalize it to other natural situations not described
by Pauli maps. Specifically, for Pauli maps we introduce an explicit
formula for the characterization of the resulting effective noise.
Beyond Pauli maps, we show that the same ideas can be applied to the
case of thermal reservoirs at arbitrary temperature. In the latter
case, we discuss how to optimize the bounds as a function of the
noise strength. We illustrate our ideas with explicit exemplary
results for several different graphs and particular decoherence
processes. The limitations of the method are also discussed.

\end{abstract}

\maketitle

%%%%%%%%%%%%%%%% Intro %%%%%%%%%%%%%%%%%%%%%%%
\section{Introduction}
\label{sec:level1} Graph states \cite{graph_review} constitute an
important family of genuine multiparticle-entangled states  with
several applications in quantum information. The most popular
example of these are arguably the cluster states, which have been
identified as a crucial resource for universal one-way measurement-based
quantum computation \cite{Brie_review,RausBrie}. Other members of
this family were also proven to be potential resources very interesting tasks, as codewords
for quantum error correction \cite{SchWer}, to implement secure
quantum communication \cite{DurCasBrie-ChenLo}, or to simulate some
aspects of the entanglement distribution of random states
\cite{Dahlsten}. Moreover, graph states encompass the celebrated
Greenberger-Horne-Zeilinger (GHZ) states \cite{GHZ}, whose
importance ranges  from fundamental to applied issues. GHZ states
can -- for large-dimensional systems -- be considered  as simple
models of the \textit{gedanken} Schr\"odinger-cat states, are
crucial for quantum communication protocols \cite{GHZuse}, and find
applications in quantum metrology \cite{Giovannetti} and
high-precision spectroscopy \cite{Bollinger}. All these reasons
explain the great deal of effort made both to theoretically
understand \cite{graph_review} the properties of, and to generate and
coherently manipulate, graph states in the
laboratory~\cite{ClusterExp}.

For the same reasons, it is crucial to unravel the dynamics of
graph states in realistic scenarios, where the system is unavoidably
exposed to interactions with its environment and/or experimental
imperfections. Previous studies on the robustness of graph-state
entanglement in the presence of decoherence showed that the disentanglement
times (\ie the time for which the state becomes separable) increases with the system size
\cite{Simon&Kempe,HeinDurBrie}. However the disentanglement time on
its own is known not to provide in general a faithful figure of
merit of the entanglement robustness:  although the disentanglement time can grow with the number $N$ of
particles, the amount of entanglement in a given time can decay exponentially with $N$
\cite{Aolita}. The full dynamical evolution must then be monitored
to draw any conclusions on the entanglement robustness.

A big obstacle must be overcome in the study of the entanglement
robustness in general mixed states: the direct quantification of
entanglement involves optimizations requiring computational
resources that increase exponentially with $N$. The problem thus
becomes in practice intractable even for relatively small system
sizes, not to mention the direct assessment of entanglement during
the entire noisy dynamics. All in all, some progress has been
achieved in the latter direction for some very particular cases: For
arbitrarily-large linear-cluster states under collective dephasing,
it is possible to calculate the exact value of the geometric measure
of entanglement throughout the evolution \cite{Guehne}.
Besides, bounds to the relative entropy and the global
robustness of entanglement for two-colorable graph
states \cite{graph_review} of any size under local dephasing were
obtained in Ref. \cite{Wunderlich}.

In a conceptually different approach, a framework to obtain families
of lower and upper bounds to the entanglement evolution of graph --
and graph-diagonal -- states under decoherence was introduced in Ref.
\cite{dan1}. The bounds are obtained via a calculation that involves
only the {\it boundary subsystem}, composed of the qubits lying at
the boundary of the multipartition under scrutiny. This, very often, reduces considerably the size of the matrices involved in the calculation of entanglement.
%and is therefore
%in very often considerably smaller than the whole system.
No optimization on the full system's parameter space is required
throughout. Another remarkable feature of the method is that it is
not limited to a particular entanglement quantifier but applies to
all convex (bi-or multi-partite) entanglement measures that do not
increase under local operations and classical communication (LOCC).
The latter are indeed two rather natural and general requirements
\cite{vidal00, Pleniohoro1}.

In the case of open-dynamic processes described by Pauli maps the
lower and upper bounds coincide and the method thus allows one to
calculate the exact entanglement of the noisy evolving state. Pauli
maps  encompass popular models of (independent or collective) noise,
as depolarization, phase flip, bit flip and bit-phase flip errors,
and are defined below. Moreover, one of the varieties of lower
bounds is of extremely simple calculation and -- despite less tight
-- depends only on the connectivity of the graph and not on its
total size. The latter is a very advantageous property in
situations where one wishes to assess the resistance of entanglement
with growing system size. For example, the versatility of the
formalism has very recently been demonstrated in Ref. \cite{dan2},
where it was applied to demonstrate the robustness of
thermal bound entanglement in  macroscopic many-body  systems of
spin-$1/2$ particles.

In the present paper, we elaborate on the details of the formalism
introduced in  \cite{dan1}. For Pauli maps we give an explicit
formula for the characterization of the effective noise involved in
the calculation of the bounds. Furthermore, we extend the method to
the case where each qubit is subject to the action of independent
thermal baths at arbitrary temperature. This is a crucial, realistic
type of dynamic process that is not described by Pauli maps. In all
cases, we exhaustively compare the different bounds with several
concrete examples. Finally, we discuss the main advantages and
limitations of our method in comparison with other approaches.

The article is organized as follows:
\begin{itemize}

\item Sec. \ref{sec:basic_concepts}:  here we define the notation, introduce definitions and review basic concepts required in the following sections.  In particular, graph and graph-diagonal states are defined in subsection \ref{graphdefinition} and the noise models considered are presented in subsection \ref{OpenSystDynamics}.

\item Sec. \ref{sec:results}: a detailed description of the proposed  framework is given in the context of fully general noises.
Families of lower and upper bounds for the entanglement evolution in
the particular multi-partition of interest in terms of the entanglement of the boundary subsystem alone
under an effective noise are provided.

\item Sec. \ref{sec:examples}: the developed machinery is applied to the case of noises described by arbitrary Pauli maps and by diffusion and dissipation with independent thermal reservoirs at any temperature. Exact results for the entanglement decay are obtained
for Pauli maps, whereas optimized bounds are
provided in the other cases.

\item Sec. \ref{sec:extlim}: we first discuss how the method can be extended to other initial states or decoherence processes. In particular, how  non-tight lower bounds for the entanglement evolution of \emph{any} initial state subjected to \emph{any} decoherence process can be obtained. Then we comment on the limitations of the method.

\item Sec. \ref{sec:conclusion}: we conclude the paper with a summary of the results and
their physical implications.

\end{itemize}

%%%%%%%%%%%%%%%%%%%%%%%%%%%%%%%%%%%%%%%%%%%%%%%%
%%%%%%%%%%%%%%%% BASIC CONCEPTS %%%%%%%%%%%%%%%%%%%%%%%

\section{\label{sec:basic_concepts}Basic concepts}

In this section, we define graph and graph-diagonal
states, introduce the basics of open-system dynamics and the
particular noise models used later.

%%%%%%%%%%%%%%%% Graph states %%%%%%%%%%%%%%%%%%%%%%%
\subsection{Graph and graph-diagonal states}
\label{graphdefinition} 
\begin{figure}[!htb]
\centering
\includegraphics[scale=0.3]{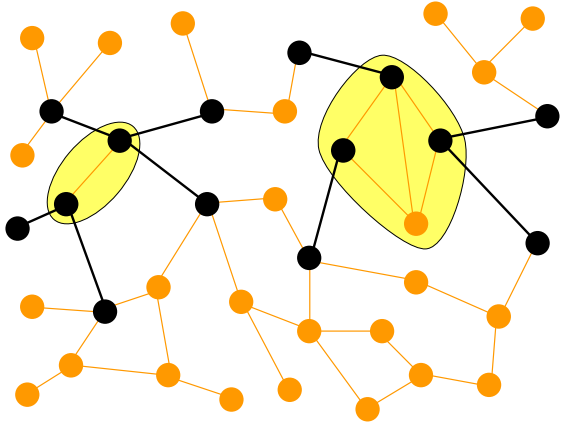}
\caption{(Color online) Mathematical graph associated to a given
physical graph state. An exemplary bipartition divides the system
into two subparts: the yellow and white regions. The edges in
black are the  \emph{boundary-crossing edges} $\mathcal{X}$ and
the nodes (also in black) connected by them are the
\emph{boundary nodes} $\mathcal{Y}$. Together they compose the
{\it boundary sub-graph} $G(\mathcal{Y},\mathcal{X})$. The
remaining vertices, painted in orange, constitute the {\it
non-boundary subsystem}.} \label{Graph_State}
\end{figure}

Qubit graph states are multiqubit quantum states
defined from mathematical graphs through the rule described below.
First, a mathematical graph
$G_{(\mathcal{V},\mathcal{C})}\equiv\{\mathcal{V},\mathcal{C}\}$ is
defined by a set $\mathcal{V}$ of $N$ vertices, or nodes, and a set
$\mathcal{C}$, of connections, or edges,  connecting each node $i$
to some other $j$. An example of such graph is illustrated in Fig.
\ref{Graph_State}. Each  vertex $i \in \mathcal{V}$ represents a
qubit in the associated physical system, and each edge $\{i,j\}\in
\mathcal{C}$ represents a unitary maximally-entangling controlled-Z
($CZ$) gate,
$CZ_{ij}=\ket{0_{i}0_{j}}\bra{0_{i}0_{j}}+\ket{0_{i}1_{j}}\bra{0_{i}1_{j}}+\ket{1_{i}0_{j}}\bra{1_{i}0_{j}}-\ket{1_{i}1_{j}}\bra{1_{i}1_{j}}$,
between the qubits $i$ and $j$ connected through the corresponding
edge. The $N$-qubit graph state
$\ket{{G_{(\mathcal{V},\mathcal{C})}}_0}$ corresponding to graph
$G_{(\mathcal{V},\mathcal{C})}$ is then operationally defined as
follows:

(i) Initialize every qubit $i$ in the superposition $\ket{+_i}=\frac{1}{\sqrt{2}}(\ket{0_i}+\ket{1_i})$, so that the joint state is in the product state
$\ket{{g_{(\mathcal{V})}}_0}\equiv\bigotimes_{i\in\mathcal{V}} \ket{+_i}$.

(ii) Then, for every connection $\{i,j\}\in\mathcal{C}$ apply the gate $CZ_{ij}$ to $\ket{{g_{(\mathcal{V})}}_0}$. That is,
\begin{equation}
\label{graphdef}
\ket{{G_{(\mathcal{V},\mathcal{C})}}_0}=\bigotimes_{\{i,j\} \in \mathcal{C}} CZ_{ij}\ket{{g_{(\mathcal{V})}}_0}.
\end{equation}

Graph state \eqref{graphdef} can also be defined in an alternative,
non-operational fashion. Associated to each node $i \in \mathcal{V}$
of a given graph $G_{(\mathcal{V},\mathcal{C})}$ we define the
operator

\be \label{stabilizer} S_{i}\equiv X_{i}\bigotimes_{j \in
\mathcal{N}_i}Z_{j}, \ee with $X_{i}$ and $Z_{j}$ the usual Pauli
operators acting respectively on qubits $i$ and $j$, and where
$\mathcal{N}_i$ denotes the set of neighbours of $i$, directly
connected to it by an edge $\{i,j\}$. Operator \eqref{stabilizer}
possess eigenvalues 1 and $-1$. It is the $i$-th generator of the
stabilizer group  and is often called for short {\it stabilizer
operator}. All $N$ stabilizer operators commute and share therefore
a common basis of eigenstates. Graph state
$\ket{{G_{(\mathcal{V},\mathcal{C})}}_0}$ in turn has the
peculiarity of being the unique common eigenstate of eigenvalue $+1$
\cite{graph_review}. In other words,

\be
\label{CommonEigen} \nonumber S_i\ket{{G_{(\mathcal{V},\mathcal{C})}}_0}=\ket{{G_{(\mathcal{V},\mathcal{C})}}_0}\  \forall\ i \in
\mathcal{V}. \ee

The other $2^N-1$ common eigenstates
$\ket{{G_{(\mathcal{V},\mathcal{C})}}_{\nu}}$ are in turn
related to  \eqref{graphdef} by a local unitary operation:
\be \label{local-unitary}
\ket{{G_{(\mathcal{V},\mathcal{C})}}_{\nu}}=\bigotimes_{i \in
\mathcal{V}}
{Z_{i}}^{\nu_i}\ket{{G_{(\mathcal{V},\mathcal{C})}}_{0}}\equiv{Z}^{\nu}\ket{{G_{(\mathcal{V},\mathcal{C})}}_{0}},
\ee
such that $S_{i}\left\vert G_{(\mathcal{V}%
,\mathcal{C}){\nu}}\right\rangle $=$\left(  -1\right)  ^{\nu_{i}%
}\left\vert G_{(\mathcal{V},\mathcal{C}){\nu}}\right\rangle $, where
$\nu$ is a multi-index representing the binary string
$\nu\equiv\nu_1\ ...\ \nu_N$, with $\nu_i=0$ or 1 $\  \forall\ i \in
\mathcal{V}$, and where the short-hand notation
${Z}^{\nu}\equiv\bigotimes_{i \in \mathcal{V}} {Z_{i}}^{\nu_i}$ has
been  introduced. Therefore, states \eqref{local-unitary} possess
all exactly the same entanglement properties and, together with
$\ket{{G_{(\mathcal{V},\mathcal{C})}}_{0}}$, define a complete
orthonormal basis of $\mathcal{H}$, called the graph-state  basis of  $\mathcal{H}$ (corresponding to
the graph ${G_{(\mathcal{V},\mathcal{C})}}$). Any state $\rho$
diagonal in such basis  is called a {\it graph-diagonal state}:
\begin{eqnarray}
\label{graphdiagonal} \rho_{GD}=\sum_{\nu} P_{\nu}
\ket{{G_{(\mathcal{V},\mathcal{C})}}_{\nu}}\bra{{G_{(\mathcal{V},\mathcal{C})}}_{\nu}},
\end{eqnarray}
where  $P_{\nu}$ is any probability distribution. Interestingly, for
any graph, any arbitrary $N$-qubit state can always be depolarized
by some separable map (defined below) into the form
\eqref{graphdiagonal} without changing its diagonal elements in the
considered graph basis \cite{Duer03Aschauer05}.

Two simple identities following from definition \eqref{stabilizer}
will be crucial for our purposes. For every eigenstate
$\ket{{G_{(\mathcal{V},\mathcal{C})}}_{\nu}}$ of $S_i$, with
eigenvalues ${s_i}_{\nu}=1$ or $-1$
\begin{eqnarray}
X_i\ket{{G_{(\mathcal{V},\mathcal{C})}}_{\nu}}&=&S_i\otimes\bigotimes_{j\in\mathcal{N}_i} Z_{j}\ket{{G_{(\mathcal{V},\mathcal{C})}}_{\nu}}
\nonumber\\
&=&{s_i}_{\nu}\bigotimes_{j\in\mathcal{N}_i} Z_{j}\ket{{G_{(\mathcal{V},\mathcal{C})}}_{\nu}},
\end{eqnarray}
where definition \eqref{stabilizer} was used,
and
\begin{eqnarray}
Y_i\ket{{G_{(\mathcal{V},\mathcal{C})}}_{\nu}}&=&(-i)Z_i.S_i\otimes\bigotimes_{j\in\mathcal{N}_i} Z_{j}\ket{{G_{(\mathcal{V},\mathcal{C})}}_{\nu}}
\nonumber\\
&=&{s_i}_{\nu} (-i)Z_{i}\otimes\bigotimes_{j\in\mathcal{N}_i} Z_{j}\ket{{G_{(\mathcal{V},\mathcal{C})}}_{\nu}}.
\end{eqnarray}
So, when applied to any pure graph -- or mixed graph-diagonal --
state, the  following operator  equivalences hold up to a global
phase:
\begin{subequations}
\label{rule}
\begin{align}
\label{sigmax} X_{i}\leftrightarrow \bigotimes_{j \in
\mathcal{N}_i}Z_{j},\\
\label{sigmay} Y_{i}\leftrightarrow Z_{i} \otimes\bigotimes_{j \in
\mathcal{N}_i}Z_{j}.
\end{align}
\end{subequations}

%%%%%%%%%%%%%%%% Noise Models %%%%%%%%%%%%%%%%%%%%%%%
\subsection{Open-system dynamics}
\label{OpenSystDynamics}

As we mentioned in Sec. \ref{sec:level1}, our ultimate goal is to study the behavior  of  graph-state entanglement in realistic dynamic scenarios where the system evolves  during a time interval $t$ according to  a generic
physical process, which can include decoherence. This
process  can always be represented by a completely-positive trace-preserving map $\Lambda$, that maps any initial state $\rho$ to the evolved one  after a time $t$, $\rho_t\equiv\Lambda(\rho)$. In turn, for every such $\Lambda$, there always exists a maximum of $D^2$ ($D=\text{dim}(\mathcal{H})$) operators $K_\mu$ such that the map is expressed in the Kraus form \cite{Nie}:
\be
\label{Kraus_rep}
\rho_t\equiv\Lambda(\rho)=\sum_\mu  K_\mu \rho
K_\mu^\dag.
\ee
Operators $K_\mu$ are called the Kraus operators, and decompose the identity operator $\mathbf{1}$ of $\mathcal{H}$ in the following manner: $\sum_\mu K_\mu^\dag K_\mu=\mathbf{1}$. Conversely, the Kraus representation encapsulates all possible physical dynamics of the system. That is, any map expressible as in \eqref{Kraus_rep} is automatically completely-positive  and trace-preserving. For our  case of interest ($N$-qubit systems), index $\mu$ runs from 0 to ${(2^N)}^2-1={4^N}-1$. For later convenience, we will represent it in base 4, decomposing it as the following multi-index: $\mu\equiv\mu_1\ ...\ \mu_N$, with $\mu_i=0$,  1, 2, or 3 $\  \forall\ i \in \mathcal{V}$.

We call $\Lambda$ a separable map with respect to some
multipartition of the system if each and all of its Kraus
operators factorize as tensor products of local operators each one
with support on only one of the subparts. For example, if we split
the qubits associated to the graph shown in Fig. \ref{Graph_State}
into a set $\mathcal{Y}$ of {\it boundary qubits} (in black) and
its complement
$\overline{\mathcal{Y}}\equiv\mathcal{V}/\mathcal{Y}$  of {\it
non-boundary qubits} (in green), $\Lambda$ is separable with
respect to this partitioning if $K_\mu\equiv
{K_\mathcal{Y}}_{\mu}\otimes {K_{\overline{\mathcal{Y}}}}_{\mu}$,
with ${K_\mathcal{Y}}_{\mu}$ and
${K_{\overline{\mathcal{Y}}}}_{\mu}$ operators acting
non-trivially only on the Hilbert spaces of the boundary and
non-boundary qubits, respectively. 

In turn, we call $\Lambda$ an {\it  independent map} with respect to
some multipartition of the system if it can be factorized as the
composition (tensor product) of individual maps acting independently
on each subpart. Otherwise, we say that $\Lambda$ is a {\it
collective map}. Examples of {\it fully independent maps} are those in
which each qubit $i$ is independently subject to its own
local noise channel $\mathcal{E}_i$. By the term {\it  independent
map} without explicit mention to any respective multipartition we
will refer throughout to fully independent maps.  In this case, the
global map $\Lambda$ factorizes completely: \be \label{individual}
\Lambda(\rho)=\mathcal{E}_1\otimes \mathcal{E}_2 \otimes \ldots
\otimes \mathcal{E}_N(\rho). \ee It is important to notice that all
independent maps are necessarily separable but a general separable
map does not need to be factorable as in \eqref{individual} and can
therefore be both, either individual or collective.

\subsubsection{Pauli maps}
\label{Pauliintro}

A crucial family of fully separable maps is that of the {\it Pauli maps}, for which every Kraus operator is proportional to a product of individual Pauli and identity operators acting on each qubit. That is, $K_\mu\equiv \sqrt{P_{(\mu_1,\ ...\ \mu_N)}}\  {\sigma_1}_{\mu_1} \otimes \ldots \otimes{\sigma_N}_{\mu_N}\equiv\sqrt{P_{\mu}}\ \sigma_{\mu}$, with ${\sigma_i}_{0}=\mathbf{1}_i$ (the identity operator on qubit $i$), ${\sigma_i}_{1}=X_i$, ${\sigma_i}_{2}=Y_i$, and ${\sigma_i}_{3}=Z_i$, and $P_{(\mu_1,\ ...\ \mu_N)}\equiv P_\mu$ any probability distribution. Popular instances are the (collective or independent) depolarization and dephasing (also called phase damping, or phase-flip) maps, and the (individual) bit-flip and bit-phase-flip channels \cite{Nie}.
For example, the independent depolarizing (D) channel describes the situation in which the qubit remains untouched with probability $1-p$, or is depolarized -- meaning that it is taken to the maximally mixed state (white noise) -- with probability $p$. It is characterized by the fully-factorable probability $P_\mu={p_1}_{\mu_1}\times\ ... \ {p_N}_{\mu_N}$, with ${p_i}_{0}=1-p$ and ${p_i}_1={p_i}_2={p_i}_3=p/3$, $\  \forall\ i \in\mathcal{V}$. The independent phase damping  (PD) channel in turn induces the complete loss of quantum coherence
with probability $p$, but without any energy (population) exchange. It is also given by a fully-factorable probability with ${p_i}_{0}=1-p/2$, ${p_i}_{1}=0={p_i}_{2}$, and  ${p_i}_{3}=p/2$, $\  \forall\ i \in\mathcal{V}$.

For later convenience, we finally recall  that  each Pauli operator ${\sigma_i}_{\mu_i}$ can be written in the following way: ${T_i}_{(u_i,v_i)}\equiv Z_i^{v_i}.X_i^{u_i}$, with $u_i$ and $v_i=0$, or 1. Indeed, notice that  ${\sigma_i}_{2v_i+|v_i+u_i|_2}={T_i}_{(u_i,v_i)}$ (up to an irrelevant phase factor for $u_i=1=v_i$), where ``$|\ |_2$" stands for modulo 2.  In this representation, called the {\it chord representation} \cite{aolita04}, the Kraus decomposition of the above-considered general Pauli map has the following Kraus operators: ${K_C}_{(U,V)}\equiv \sqrt{P_{C(u_1,v_1,\ ...\ u_N,v_N)}}\ {T_1}_{(u_1,v_1)} \otimes \ldots \otimes{T_N}_{(u_N,v_N)}\equiv\sqrt{P_{C(U,V)}}\ {T}_{(U,V)}$, where $U\equiv(u_1,\ ...\ u_N)$ and $V\equiv(v_1,\ ...\ v_N)$. The probability $P_{C(U,V)}\equiv P_{C(u_1,v_1,\ ...\ u_N,v_N)}$ in turn is related to the  original $P_{\mu}$ by $P_{C(u_1,v_1,\ ...\ u_N,v_N)}\equiv P_{(2 v_1+|v_1+u_1|_2,\ ...\ ,2 v_N+|v_N+u_N|_2))}$.

\subsubsection{Independent thermal baths}
\label{Thermaldef}

\par An important example of a non-Pauli, independent map is the generalized amplitude-damping channel (GAD) \cite{Nie}. It represents energy diffusion and dissipation with a thermal bath into which each qubit is individually immersed. Its Kraus representation is
\begin{subequations}\label{Kraus_AmpDamp}
\be {K_i}_{\mu_i=0}\equiv\sqrt{\frac{\overline{n}+1}{2\overline{n}+1}}(\ket{0_i}\bra{0_i}+\sqrt{1-p}\ket{1_i}\bra{1_i}),\ee
\be{K_i}_{\mu_i=1}\equiv\sqrt{\frac{\overline{n}+1}{2\overline{n}+1}p}\ket{0_i}\bra{1_i},\ee
\be{K_i}_{\mu_i=2}\equiv\sqrt{\frac{\overline{n}}{2\overline{n}+1}}(\sqrt{1-p}\ket{0_i}\bra{0_i}+\ket{1_i}\bra{1_i}),\ee and
\be{K_i}_{\mu_i=3}\equiv\sqrt{\frac{\overline{n}}{2\overline{n}+1}p}\ket{1_i}\bra{0_i}.\ee
\end{subequations}
Here $\overline{n}$ is the average number
of quanta in the thermal bath, $p\equiv p(t)\equiv
1-e^{-\frac{1}{2}\gamma(2\overline{n}+1)t}$ is the probability of
the qubit exchanging a quantum with the bath after a  time $t$, and
$\gamma$ is the zero-temperature dissipation rate.
Channel GAD is actually the extension to finite temperature of
the purely dissipative amplitude damping (AD) channel, which is
obtainen from GAD in the zero-temperature limit $\overline{n}=0$. In the opposite extreme, the purely diffusive case is
obtained from GAD in the composite limit $\overline{n}\rightarrow\infty$, $\gamma\rightarrow0$, and
$\overline{n}\gamma=\Gamma$, where $\Gamma$ is the diffusion
constant. Note that in the purely-diffusive limit, channel GAD becomes a Pauli channel, with defining individual probabilities ${p_i}_{0}=\frac{1}{2}(1-p/2+\sqrt{1-p})$, ${p_i}_{1}=\frac{p}{4}={p_i}_{2}$, and  ${p_i}_{3}=\frac{1}{2}(1-p/2-\sqrt{1-p})$, $\  \forall\ i \in\mathcal{V}$.

\par Finally, the probability $p$ in channels D, PD and GAD above can be interpreted as a convenient parametrization of time, where $p=0$ refers
to the initial time 0 and $p=1$ refers to the asymptotic limit $t\rightarrow\infty$.

%%%%%%%%%%%%%%%% FORMALISM %%%%%%%%%%%%%%%%%%%%%%%
%%%%%%%%%%%%%%%%%%%%%%%%%%%%%%%%%%%%%%%%%%%%%

\section{\label{sec:results} Evolution of  graph-state entanglement under generic noise}
As mentioned before, the direct calculation of the entanglement in
arbitrary mixed states is a task exponentially hard in the system's
size \cite{Pleniohoro1}. In this section, we elaborate in detail a
formalism that dramatically simplifies this task for graph -- or
graph-diagonal -- states undergoing a noisy evolution in a fully
general context. Along the way, we also describe carefully which
requirements an arbitrary noisy map has to satisfy so that the
formalism can be applied.
%%%%%%%%%%%%%%%% Formalism %%%%%%%%%%%%%%%%%%%%%%%
\subsection{\label{sec:general} The general idea}
Consider a system initially in graph state \eqref{graphdef} that
evolves during a time $t$ according to the general map
\eqref{Kraus_rep} towards the evolved state
\begin{eqnarray}
\label{evolved}
\rho_t\equiv\Lambda(\ket{{G_{(\mathcal{V},\mathcal{C})}}_0})=\sum_{\mu}
K_\mu
\ket{{G_{(\mathcal{V},\mathcal{C})}}_0}\bra{{G_{(\mathcal{V},\mathcal{C})}}_0}
K_\mu^\dag.
\end{eqnarray}
We would like to follow the entanglement $E(\rho_t)$  of $\rho_t$ during its entire evolution. Here, $E$ is any convex entanglement monotone \cite{vidal00, Pleniohoro1} that quantifies the entanglement content in some given multi-partition of the system. An example of such multi-partition is displayed in Fig. \ref{Graph_State}, where the associated graph is split into two subsets, painted respectively in yellow and white in the figure. The
edges that connect vertices at different subsets are called the  \emph{boundary-crossing edges} and are painted in black in the figure. We call the set of all the boundary-crossing edges $\mathcal{X}\subseteq\mathcal{C}$, and its complement $\overline{\mathcal{X}}\equiv\mathcal{C}/\mathcal{X}$ the set of all non-boundary-crossing edges. All the qubits associated to vertices connected by any edge in $\mathcal{X}$ constitute the set $\mathcal{Y}\subseteq\mathcal{V}$ of \emph{boundary qubits} (or \emph{boundary subsystem}), and its complement $\overline{\mathcal{Y}}\equiv\mathcal{V}/\mathcal{Y}$ is the non-boundary qubit set. We refer to  $G_{(\mathcal{Y},\mathcal{X})}$ as the {\it boundary sub-graph}.

We can use this classification and the operational definition  \eqref{graphdef} to write the initial graph state as
\be\label{explicitCZ}
\ket{{G_{(\mathcal{V},\mathcal{C})}}_0}=\bigotimes_{\{i,j\} \in \overline{\mathcal{X}}} CZ_{ij}\ket{{G_{(\mathcal{Y},\mathcal{X})}}_0}\otimes\ket{{g_{(\overline{\mathcal{Y}})}}_0},\ee
where
$\ket{{g_{(\overline{\mathcal{Y}})}}_0}\equiv\bigotimes_{i\in\overline{\mathcal{Y}}} \ket{+_i}$. In other words, we explicitly factor all the $CZ$ gates corresponding to non-boundary qubits out.

Consider now the application of some Kraus operator $K_\mu$ of a general map on graph state \eqref{explicitCZ}: $K_\mu\bigotimes_{\{i,j\} \in \overline{\mathcal{X}}} CZ_{ij}\ket{{G_{(\mathcal{Y},\mathcal{X})}}_0}\otimes\ket{{g_{(\overline{\mathcal{Y}})}}_0}$. The latter can  always be written as $\bigotimes_{\{i,j\} \in \overline{\mathcal{X}}} CZ_{ij} {\tilde{K}}_\mu\ket{{G_{(\mathcal{Y},\mathcal{X})}}_0}\otimes\ket{{g_{(\overline{\mathcal{Y}})}}_0}$, with
\be
\label{commutation}
{\tilde{K}}_\mu=\bigotimes_{\{i,j\}\in
\overline{\mathcal{X}}}CZ_{ij}\ K_{\mu}\bigotimes_{\{i',j'\}\in \overline{\mathcal{X}}}CZ_{i'j'},\ \forall\ \mu ,
\ee
Now, consider every map $\Lambda$ such that transformation rule \eqref{commutation} yields, for each $\mu$, modified Kraus operators
of the form
\begin{equation}
\label{bisepKraus}
\tilde{K}_{\mu}={{\tilde{K}}_{\mathcal{Y}\gamma}}\otimes{{\tilde{K}}_{\overline{\mathcal{Y}}\omega}},
\end{equation}
where ${{\tilde{K}}_{\mathcal{Y}\gamma}}$ and ${{\tilde{K}}_{\overline{\mathcal{Y}}\omega}}$ are normalized modified Kraus operators acting non-trivially only on the boundary and non-boundary qubits, respectively. In the last, $\gamma=\{\mu_i,\ \ i\in\mathcal{Y}\}$ and $\omega=\{\mu_i,\ \ i\in\overline{\mathcal{Y}}\}$ are multi-indices labeling respectively the alternatives for the boundary and non-boundary subsystems, being $\mu_i$ in turn the individual base-4 indices introduced after Eq. \eqref{Kraus_rep}. The modified map $\tilde{\Lambda}$, composed of Kraus operators $\tilde{K}_{\mu}$ is then clearly  bi-separable with respect to the bi-partition ``boundary $/$ non-boundary". For all such maps the calculation of  $E(\rho_t)$  can be drastically simplified, as we see in what follows.

In these cases, the evolved state \eqref{evolved} can be writen as
\begin{eqnarray}
\label{evolvedcommuted}
\rho_t\equiv\Lambda(\ket{{G_{(\mathcal{V},\mathcal{C})}}_0})=
\bigotimes_{\{i,j\}\in \overline{\mathcal{X}}}CZ_{ij}\ \tilde{\rho}_t\bigotimes_{\{k,l\}\in \overline{\mathcal{X}}}CZ_{kl}.
\end{eqnarray}
with
\begin{widetext}
\begin{eqnarray}
\label{evolvedcommutedtilde}
\nonumber\tilde{\rho}_t=\tilde{\Lambda}(\ket{{G_{(\mathcal{Y},\mathcal{X})}}_0}\otimes\ket{{g_{(\overline{\mathcal{Y}})}}_0})=
\sum_{\mu} {\tilde{K}}_{\mathcal{Y}\gamma(\mu)}\ket{{G_{(\mathcal{Y},\mathcal{X})}}_0}\bra{{G_{(\mathcal{Y},\mathcal{X})}}_0}
{\tilde{K}}^\dag_{\mathcal{Y}\gamma(\mu)}\otimes{\tilde{K}}_{\overline{\mathcal{Y}}\omega(\mu)}\ket{{g_{(\overline{\mathcal{Y}})}}_0}\bra{{g_{(\overline{\mathcal{Y}})}}_0}{\tilde{K}}^\dag_{\overline{\mathcal{Y}}\omega(\mu)}\\
=\sum_{\omega}
{\tilde{K}}_{\overline{\mathcal{Y}}\omega}\ket{{g_{(\overline{\mathcal{Y}})}}_0}\bra{{g_{(\overline{\mathcal{Y}})}}_0}{\tilde{K}}^\dag_{\overline{\mathcal{Y}}\omega}\otimes\sum_{\gamma}
{\tilde{K}}_{\mathcal{Y}(\gamma|\omega)}\ket{{G_{(\mathcal{Y},\mathcal{X})}}_0}\bra{{G_{(\mathcal{Y},\mathcal{X})}}_0}
{\tilde{K}}^\dag_{\mathcal{Y}(\gamma|\omega)},
\end{eqnarray}
\end{widetext}
where ${\tilde{K}}_{\mathcal{Y}(\gamma|\omega)}$ is the $\gamma$-th modified Kraus operator on the boundary subsystem given that ${\tilde{K}}_{\overline{\mathcal{Y}}\omega}$ has been applied to the non-boundary one. Recall that both  $\gamma\equiv\gamma(\mu)$ and $\omega\equiv\omega(\mu)$ come from the same single multi-index $\mu$ and are therefore in general not independent on one another. In the second equality of \eqref{evolvedcommutedtilde} we have chosen to treat $\omega$ as an independent variable for the summation and make $\gamma$ explicitly depend on $\omega$. This can always be done and will be convenient for our purposes.

The crucial observation now is that the $CZ$ operators explicitly factored out in the evolved state
\eqref{evolvedcommuted} correspond to non-boundary-crossing edges. Thus,  they act as \emph{local unitary operations} with respect to
the multi-partition of interest. For this reason, and since local unitary operations do not change the entanglement content of any state, the equivalence
\begin{equation}
\label{equivalence}
E(\rho_t)= E(\tilde{\rho}_t)
\end{equation}
holds.

\par
\begin{figure}[!htb]
\centering
\includegraphics[scale=0.3]{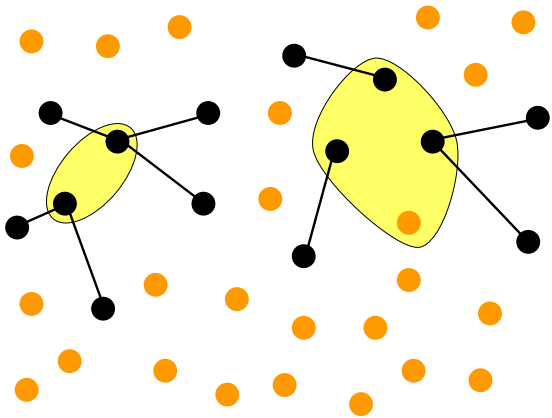}
\caption{(Color online) Same graph as in Fig. \ref{Graph_State}
but where all non-boundary-crossing edges have been erased,
representing the fact that the boundary and non-boundary
subsystems are fully unentangled. The entanglement in the whole
system is obtained via a calculation involving only the smaller
boundary subsystem.} \label{Graph_State_factored}
\end{figure}

 In the forthcoming subsections we see how, by exploiting this equivalence in different noise scenarios, the computational effort required for a reliable estimation (and in some cases, an exact calculation) of $E(\rho_t)$ can be considerably reduced. The main idea behind this reduction lies on the fact that, whereas in general state \eqref{evolved} the entanglement can be distributed among all particles in the graph, in state \eqref{evolvedcommutedtilde} the boundary and non-boundary subsystems are explicitly in a separable state. All the entanglement in the multi-partition of interest is therefore  localized exclusively in the boundary subgraph. The situation is graphically represented in Fig. \ref{Graph_State_factored}, where the same graph as in Fig. \ref{Graph_State} is plotted but with all its non-boundary-crossing edges erased.

More precisely, the general approach consists of obtaining lower and upper bounds on $E(\rho_t)$ by bounding the entanglement of state \eqref{evolvedcommutedtilde} from above and below as explained in what follows.

\subsubsection{Lower bounds to the entanglement evolution}

The property of LOCC monotonicity of $E$, which  means that the average entanglement cannot grow during an LOCC process \cite{vidal00}, allows us to derive lower-bounds on $E(\tilde{\rho}_t)$. The ones we consider can be obtained by the following generic procedure:
\begin{itemize}
\item[{\it (i)}] after bringing the studied state into the form \eqref{evolvedcommutedtilde}, apply some local  general measurement $M=\{M_{\omega'}\}$, with measurement elements $M_{\omega}$, on the non-boundary subsystem $\overline{\mathcal{Y}}$;
\item[{\it (ii)}] for each measurement outcome $\omega$ trace out the measured non-boundary subsystem;
\item[{\it (iii)}]  and finally, calculate the mean entanglement in the resulting state of the boundary subsystem $\mathcal{Y}$, averaged over all outcomes $\omega$.
\end{itemize}
Since this procedure constitutes an LOCC with respect to the multipartition under scrutiny, the latter average entanglement can only be smaller than, or equal to, that of the initial state, \ie:
\begin{widetext}
\begin{eqnarray}
\label{lower}
E(\rho_t)= E(\tilde{\rho}_t)\geq\sum_{\omega}P_{\omega}E\Big(\sum_{\omega'}\frac{1}{P_{\omega}}\bra{{g_{(\overline{\mathcal{Y}})}}_0}{\tilde{K}}^\dag_{\overline{\mathcal{Y}}\omega'}M^\dag_{\omega}.M_{\omega}{\tilde{K}}_{\overline{\mathcal{Y}}\omega'}\ket{{g_{(\overline{\mathcal{Y}})}}_0}\sum_{\gamma} {\tilde{K}}_{\mathcal{Y}(\gamma|\omega')}\ket{{G_{(\mathcal{Y},\mathcal{X})}}_0}\bra{{G_{(\mathcal{Y},\mathcal{X})}}_0}
{\tilde{K}}^\dag_{\mathcal{Y}(\gamma|\omega')}\Big).
\end{eqnarray}
\end{widetext}
with $P_{\omega}\equiv\sum_{\omega'}\bra{{g_{(\overline{\mathcal{Y}})}}_0}{\tilde{K}}^\dag_{\overline{\mathcal{Y}}\omega'}M^\dag_{\omega}.M_{\omega}{\tilde{K}}_{\overline{\mathcal{Y}}\omega'}\ket{{g_{(\overline{\mathcal{Y}})}}_0}$ being the probability of outcome $\omega$.

Notice that if the states $\{{\tilde{K}}_{\overline{\mathcal{Y}}\omega'}\ket{{g_{(\overline{\mathcal{Y}})}}_0}\}$ of the non-boundary subsystem happen to be orthogonal, then there exists an optimal measurement $M=\{M_{\omega}\equiv\frac{{\tilde{K}}_{\overline{\mathcal{Y}}\omega}\ket{{g_{(\overline{\mathcal{Y}})}}_0}\bra{{g_{(\overline{\mathcal{Y}})}}_0}{\tilde{K}}^\dag_{\overline{\mathcal{Y}}\omega}}{\bra{{g_{(\overline{\mathcal{Y}})}}_0}{\tilde{K}}^\dag_{\overline{\mathcal{Y}}\omega}{\tilde{K}}_{\overline{\mathcal{Y}}\omega}\ket{{g_{(\overline{\mathcal{Y}})}}_0}}\}$ that can distinguish them unambiguously, so that $\bra{{g_{(\overline{\mathcal{Y}})}}_0}{\tilde{K}}^\dag_{\overline{\mathcal{Y}}\omega'}M^\dag_{\omega}.M_{\omega}{\tilde{K}}_{\overline{\mathcal{Y}}\omega'}\ket{{g_{(\overline{\mathcal{Y}})}}_0}=\delta_{\omega,\omega'}\times\bra{{g_{(\overline{\mathcal{Y}})}}_0}{\tilde{K}}^\dag_{\overline{\mathcal{Y}}\omega}{\tilde{K}}_{\overline{\mathcal{Y}}\omega}\ket{{g_{(\overline{\mathcal{Y}})}}_0}$ and $P_{\omega}=\bra{{g_{(\overline{\mathcal{Y}})}}_0}{\tilde{K}}^\dag_{\overline{\mathcal{Y}}\omega}{\tilde{K}}_{\overline{\mathcal{Y}}\omega}\ket{{g_{(\overline{\mathcal{Y}})}}_0}$. In these cases an optimal lower bound is achieved as
\begin{eqnarray}
\label{loweroptimal}
\nonumber E(\rho_t)\geq\\
\sum_{\omega} P_{\omega} E\big(\sum_{\gamma} {\tilde{K}}_{\mathcal{Y}(\gamma|\omega)}\ket{{G_{(\mathcal{Y},\mathcal{X})}}_0}\bra{{G_{(\mathcal{Y},\mathcal{X})}}_0}
{\tilde{K}}^\dag_{\mathcal{Y}(\gamma|\omega)}\big).
\end{eqnarray}
Full distinguishability of the states in the non-boundary subsystem allows to reduce the mixing in the remaining boundary subsystem. In other words, the measurement outcome $\omega$ works as a perfect {\it flag} that marks which sub-ensemble of states of the boundary subsystem, from all those  present in
mixture \eqref{evolvedcommutedtilde}, corresponds indeed to the obtained outcome.

In the opposite extreme, when states $\{{\tilde{K}}_{\overline{\mathcal{Y}}\omega'}\ket{{g_{(\overline{\mathcal{Y}})}}_0}\}$ are all equal, no flagging information can be obtained via any measurement. In this case, the resulting bound is always equal to that obtained had we not made any measurement at all, but just directly taken the partial trace over $\overline{\mathcal{Y}}$ from \eqref{evolvedcommutedtilde}:
\begin{eqnarray}
\label{lowernonoptimal}
\nonumber E(\rho_t)\geq\\
E\big(\frac{1}{2^{|\overline{\mathcal{Y}}|}}\sum_{\omega,\gamma}{\tilde{K}}_{\mathcal{Y}(\gamma|\omega)}\ket{{G_{(\mathcal{Y},\mathcal{X})}}_0}\bra{{G_{(\mathcal{Y},\mathcal{X})}}_0}
{\tilde{K}}^\dag_{\mathcal{Y}(\gamma|\omega)}\big).
\end{eqnarray}
where $|\overline{\mathcal{Y}}|$ stands for the number of non-boundary qubits and full mixing over variable $\omega$ takes place now.

Henceforth we refer to lower bound \eqref{lowernonoptimal} as the {\it lowest lower bound} (LLB). As its name suggests, its tightness is far from the optimal one given by \eqref{loweroptimal}. However, as we will see in the forthcoming subsections, due to the partial tracing, it typically does not depend  on the total system's size but just on  the boundary subsystem's.

This constitutes an appealing, useful property, for it allows one to draw generic conclusions about the robustness of entanglement in certain partitions of graph states, irrespective of their number of constituent particles (see examples in Sec. \ref{GAD}).

%entanglement of all noisy graph states of arbitrary size which share in common the boundary subgraph. ({\bf Perhaps cite here our latest paper in New Journal of Phys...})

\subsubsection{Upper bounds to the entanglement evolution}

On the other hand, we consider upper-bounds on $E(\rho_t)$ based on the property of convexity of $E$, which essentially means that the entanglement of the convex sum is lower than, or equal to, the convex sum of the entanglements \cite{vidal00, Pleniohoro1}. From \eqref{evolvedcommutedtilde}, the latter implies that
\begin{widetext}
\be
E(\rho_t)= E(\tilde{\rho}_t)\leq\sum_{\omega}P_{\omega}E\big(\frac{1}{P_{\omega}}{\tilde{K}}_{\overline{\mathcal{Y}}\omega}\ket{{g_{(\overline{\mathcal{Y}})}}_0}\bra{{g_{(\overline{\mathcal{Y}})}}_0}{\tilde{K}}^\dag_{\overline{\mathcal{Y}}\omega}\otimes\sum_{\gamma}{\tilde{K}}_{\mathcal{Y}(\gamma|\omega)}\ket{{G_{(\mathcal{Y},\mathcal{X})}}_0}\bra{{G_{(\mathcal{Y},\mathcal{X})}}_0}
{\tilde{K}}^\dag_{\mathcal{Y}(\gamma|\omega)}\big),
\ee
\end{widetext}
where, once again, $P_{\omega}=\bra{{g_{(\overline{\mathcal{Y}})}}_0}{\tilde{K}}^\dag_{\overline{\mathcal{Y}}\omega}{\tilde{K}}_{\overline{\mathcal{Y}}\omega}\ket{{g_{(\overline{\mathcal{Y}})}}_0}$. In each term of the last summation the boundary and non-boundary subsystems inside the brackets are in a {\it product state}. Therefore, as for what the multi-partition of interest concerns, the non-boudary subsystem works as a locally-added ancila (in a state $\frac{1}{P_{\omega}}{\tilde{K}}_{\overline{\mathcal{Y}}\omega}\ket{{g_{(\overline{\mathcal{Y}})}}_0}\bra{{g_{(\overline{\mathcal{Y}})}}_0}{\tilde{K}}^\dag_{\overline{\mathcal{Y}}\omega}$) and consequently does not have any influence on the amount of entanglement. This leads to the generic upper bound
\begin{eqnarray}
\label{uppernonoptimal}
\nonumber E(\rho_t)\leq\\
\sum_{\omega}P_{\omega}E\big(\sum_{\gamma} {\tilde{K}}_{\mathcal{Y}(\gamma|\omega)}\ket{{G_{(\mathcal{Y},\mathcal{X})}}_0}\bra{{G_{(\mathcal{Y},\mathcal{X})}}_0}
{\tilde{K}}^\dag_{\mathcal{Y}(\gamma|\omega)}\big).
\end{eqnarray}

\subsubsection{Exact entanglement}
\label{ExactEnt}

\par Notice that upper bound \eqref{uppernonoptimal} and optimal lower bound \eqref{loweroptimal} coincide. This means that, in the above-mentioned case when states $\{{\tilde{K}}_{\overline{\mathcal{Y}}\omega}\ket{{g_{(\overline{\mathcal{Y}})}}_0}\}$ are orthogonal, these coincident bounds yield actually the exact value  of $E(\rho_t)$:
\begin{eqnarray}
\label{exact}
\nonumber E(\rho_t)=\\
\sum_{\omega}P_{\omega}E\big(\sum_{\gamma} {\tilde{K}}_{\mathcal{Y}(\gamma|\omega)}\ket{{G_{(\mathcal{Y},\mathcal{X})}}_0}\bra{{G_{(\mathcal{Y},\mathcal{X})}}_0}
{\tilde{K}}^\dag_{\mathcal{Y}(\gamma|\omega)}\big).
\end{eqnarray}

\par Expression \eqref{exact}  is still not an analytic closed formula for the exact entanglement of $\rho_t$, but reduces its calculation to that of the average entanglement over an ensemble of states of the boundary subsystem alone. More in detail, a brute-force calculation of $E(\rho_t)$ would require in general a convex optimization over the entire ${(2^N)}^2$-complex-parameter space. Through Eq. \eqref{exact} in turn such calculation is reduced to
that of the average entanglement over a sample of $2^{|\overline{\mathcal{Y}}|}$  states (one for each
$\omega$) of  $|\mathcal{Y}|$ qubits, being $|\mathcal{Y}|$  the number of boundary qubits. The latter involves at most $2^{|\overline{\mathcal{Y}}|}$ independent optimizations over a ${(2^{|\mathcal{Y}|})}^2$-complex-parameter space. This, from the point of view of  computational memory required, accounts for a reduction of resources by a factor of ${(2^{|\overline{\mathcal{Y}}|})}^2$. Alternatively, when computational memory is not a major restriction -- for example if large classical-computer clusters are at hand --, one can take advantage of the fact that the $|\overline{\mathcal{Y}}|$ required optimizations in \eqref{exact} are independent and therefore the calculation comes readily perfectly-suited for parallel computing. In this case, it is in the required computational time where an $|\overline{\mathcal{Y}}|$-exponentially large speed-up is gained.

\par In the cases where states $\{{\tilde{K}}_{\overline{\mathcal{Y}}\omega}\ket{{g_{(\overline{\mathcal{Y}})}}_0}\}$ are not orthogonal and the upper and lower bounds do not coincide, expressions  \eqref{uppernonoptimal} and \eqref{lower} still yield highly non-trivial upper and lower bounds, respectively, as we  discuss in Sec. \ref{GAD}.

Finally, it is important to stress that all the bounds  derived here
are general in the sense that they hold for  any function fulfilling
the {\it fundamental  properties of convexity and monotonicity under
LOCC processes}. This class includes genuine multipartite
entanglement measures, as well as several quantities designed to
quantify the usefulness of quantum states in the fulfillment of some
given task for quantum-information processing or communication
\cite{Pleniohoro1}.
%%%%%%%%%%%%%%%% Formalism %%%%%%%%%%%%%%%%%%%%%%%

\section{\label{sec:examples} Graph states under Pauli maps or thermal reservoirs}

In the present section we apply the ideas of the previous section to some important concrete examples of noise processes. This shows how the method is helpful in the entanglement calculation for systems in natural, dynamic physical scenarios. We first discuss the case of Pauli maps and then the generalized amplitude damping channel (thermal reservoir). Explicit calculations for noisy graph states composed of up to fourteen qubits are presented as examples.

\subsection{\label{sec:pauli}  Pauli maps on graph states}
Pauli maps defined in Sec. \ref{OpenSystDynamics} provide the most important and general subfamily of noise types for which expression \eqref{exact}  for the exact entanglement of the evolved state applies. In this case, every $X_i$ or $Y_i$ Pauli matrix in the map's Kraus operators is systematically substituted by products of $Z_i$ and $\mathbf{1}_i$ according to rules \eqref{rule}.  The resulting map $\tilde{\Lambda}$ defined in this way automatically  commutes with any $CZ$ gate and is fully separable, so that condition \eqref{bisepKraus} is trivially satisfied. Since for every qubit in the system four orthogonal single qubit operators are mapped into products of just two, several different Kraus operators of the original map contribute to the same Kraus operator of the modified one. This allows us to simplify the notation going from indices $\mu_i$, which run over $4$ possible values each, to modified indices $\tilde{\mu}_i$ having only two different alternatives. In fact, the original operators $K_\mu$ give rise to only $2^N$  modified ones of the form
\be \label{Kraus_generic}
\tilde{K}_{\tilde{\mu}}=\sqrt{\tilde{P}_{\tilde{\mu}}} Z_1^{{\tilde{\mu}}_1}
\otimes Z_2^{{\tilde{\mu}}_2} \otimes \dots \otimes Z_N^{{\tilde{\mu}}_N}\equiv\sqrt{\tilde{P}_{\tilde{\mu}}}Z^{{\tilde{\mu}}}
\ee where multi-index $\tilde{\mu}$ stands for the binary string $\tilde{\mu}=\tilde{\mu}_1\ \dots\ \tilde{\mu}_N$, with $\mu_i=0$ or 1, $\forall i\in\mathcal{V}$. Probability $\tilde{P}_{\tilde{\mu}}$ is given simply by the summation of all $P_\mu$ in the original Pauli map over all the different events $\mu$ for which $\sigma_{\mu}$ yields -- via rules \eqref{rule} -- the same modified operator $Z^{{\tilde{\mu}}}$ in \eqref{Kraus_generic}.

\par To compute the latter modified probability we move to the chord notation  \cite{aolita04}, mentioned at the end of Sec. \ref{Pauliintro}. Indeed, under transformation \eqref{rule}, we have that  ${T_i}_{(u_i,v_i)}\rightarrow  Z_i^{{v_i}_1}\otimes\bigotimes_{j \in
\mathcal{N}_i}Z_{j}^{u_i}$, so that ${T}_{(U,V)}\equiv{T_1}_{(u_1,v_1)}\otimes\ ... \ {T_N}_{(u_N,v_N)}\rightarrow Z_1^{v_1+\sum_{j \in
\mathcal{N}_1}u_j}\otimes  \dots \otimes Z_N^{v_N+\sum_{j \in
\mathcal{N}_N}u_j}$. The latter coincides with  $Z^{{\tilde{\mu}}}$ every time $\tilde{\mu}_i=|v_i+\sum_{j \in
\mathcal{N}_i}u_j|_2,\  \forall\ i \in\mathcal{V}$. Thus,  in this representation, the modified probability $\tilde{P}_{C\tilde{\mu}}$ is obtained from the defining probability $P_{C(U,V)}$ in the original map by the explicit formula
\begin{equation}
\tilde{P}_{C\tilde{\mu}}\equiv\sum_{U} P_{C(u_1,|\tilde{\mu}_1-\sum_{j \in
\mathcal{N}_1}u_j|_2,\ ... \ ,u_N,|\tilde{\mu}_1-\sum_{j \in
\mathcal{N}_N}u_j|_2)}.
\end{equation}

The modified Kraus operators \eqref{Kraus_generic}  in turn  are fully
separable; thus, as said, they trivially satisfy factorization condition \eqref{bisepKraus}. We can express them as $\tilde{K}_{\tilde{\mu}}={\tilde{K}}_{\mathcal{Y}\tilde{\gamma}}\otimes{\tilde{K}}_{\overline{\mathcal{Y}}\tilde{\omega}}$, with
\begin{equation}
\label{Kraus_separable}
{\tilde{K}}_{\mathcal{Y}\tilde{\gamma}}\equiv{\tilde{K}}_{\mathcal{Y}(\tilde{\gamma}|\tilde{\omega})}=\sqrt{\tilde{P}_{(\tilde{\gamma}|\tilde{\omega})}}Z^{{\tilde{\gamma}}}%\equiv\tilde{P}_{\tilde{\gamma}}\bigotimes_{ i\in\mathcal{Y}}Z_i^{{\tilde{\mu}}_i}\
\text{ and }\ {\tilde{K}}_{\overline{\mathcal{Y}}\tilde{\omega}}=\sqrt{\tilde{P}_{\tilde{\omega}}}Z^{{\tilde{\omega}}}%\equiv\tilde{P}_{\tilde{\omega}}\bigotimes_{ i\in\overline{\mathcal{Y}}}Z_i^{{\tilde{\mu}}_i}
.
\end{equation}
The new  multi-indices are   $\tilde{\gamma}=\{\tilde{\mu}_i,\ i\in\mathcal{Y}\}$ and $\tilde{\omega}=\{\tilde{\mu}_i,\ i\in\overline{\mathcal{Y}}\}$, and the corresponding probabilities  satisfy $\tilde{P}_{(\tilde{\gamma}|\tilde{\omega})}\tilde{P}_{\tilde{\omega}}\equiv \tilde{P}_{\tilde{\mu}}$.

\begin{figure}
\centering
\includegraphics[scale=0.3]{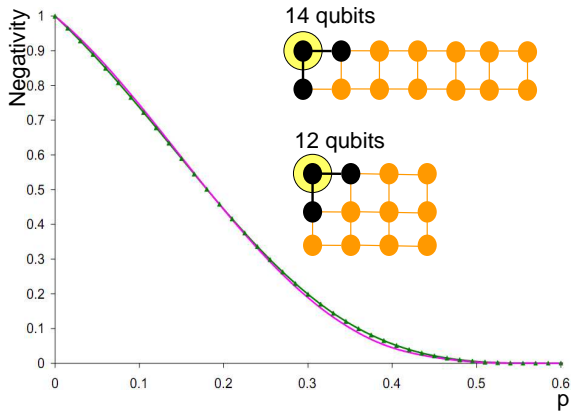}
\caption{(Color online) Negativity vs $p$ for 14-qubit
(green triangles) and 12-qubit (pink solid curve) cluster states
 undergoing independent depolarizing noise, and for the bipartition shown in inset.
Parameter $p$ can be thought as a parametrization of time (see
text).} \label{12qubits_negativity}
\end{figure}

The states $\{{\tilde{K}}_{\overline{\mathcal{Y}}\tilde{\omega}'}\ket{{g_{(\overline{\mathcal{Y}})}}_0}=\sqrt{\tilde{P}_{\tilde{\omega}'}}Z^{{\tilde{\omega}'}}\ket{+_i}=\sqrt{\tilde{P}_{\tilde{\omega}'}}\bigotimes_{ i\in\overline{\mathcal{Y}}}\frac{1}{\sqrt{2}}(\ket{0_i}+(-1)^{{\tilde{\mu}}_i}\ket{1_i})\equiv\sqrt{\tilde{P}_{\tilde{\omega}'}}\ket{{g_{(\overline{\mathcal{Y}})}}_{\tilde{\omega}'}}\}$  are trivially checked to be all orthogonal. Thus, they provide perfect flags that mark each sub-ensemble in  the boundary subsystem's ensemble. The perfect flags are revealed  by local measurements on the non-boundary qubits in the product basis $\{\ket{{g_{(\overline{\mathcal{Y}})}}_{\tilde{\omega}}}\}$. Therefore, for Pauli maps the exact entanglement $E(\rho_t)$ can be calculated by expression \eqref{exact}, which, in terms of binary indeces $\tilde{\gamma}$ and $\tilde{\omega}$, and using graph-state relationship \eqref{local-unitary}, can be finally expressed as
\begin{eqnarray}
\label{graphdiagonalexact}
E(\rho_t)=\sum_{\tilde{\omega}} \tilde{P}_{\tilde{\omega}}
E\big(\sum_{\tilde{\gamma}} \tilde{P}_{(\tilde{\gamma}|\tilde{\omega})}\ket{{G_{(\mathcal{Y},\mathcal{X})}}_{\tilde{\gamma}}}\bra{{G_{(\mathcal{Y},\mathcal{X})}}_{\tilde{\gamma}}}\big),
\end{eqnarray}
In Fig. \ref{12qubits_negativity} we have plotted the
bipartite entanglement of the exemplary bipartition of one qubit versus the rest shown in its inset for  fourteen and  twelve qubit
graph states evolving under individual depolarization. This map, as said before, is characterized by the one-qubit Kraus operators $\sqrt{1-p}\mathbf{1}, \sqrt{p/3}X,\sqrt{p/3}Y$, and $\sqrt{p/3}Z$. The parameter $p$ ($0\leq p\leq1$) refers to the probability that the map has acted: for $p=0$ the state is left untouched and for $p=1$ it is completely depolarized. Once more, $p$ can be also set as a parametrization of time: $p=0$ referring to the initial time (when nothing has occurred) and $p=1$ referring to the asymptotic time $t\rightarrow \infty$ (when the system reaches its final steady state).

%%%%%%%%%%%%%%%%%%%%%%%%%%%%%%   Insert a nine qubit cluster state    %%%%%%%%%%%%%%%%%%%%%%%%%%%%%%

As the quantifier of entanglement, we choose the negativity \cite{VidWer}, defined as the absolute value of the sum of the
negative eigenvalues of the  density matrix partially transposed with respect to the considered bipartition. It is a convex entanglement monotone that in general fails to quantify entanglement of some entangled
states  -- those ones with positive partial transposition (PPT) -- in
dimensions higher than six \cite{Pleniohoro1}.
However, since its calculation does not require optimizations but just matrix diagonalizations, it is very well-suited for a simple illustration of our ideas.

We emphasize that, for the graph used in Fig. \ref{12qubits_negativity}, a brute-force calculation would involve diagonalizing a
$2^{14}\times2^{14}=16384\times16384$ density matrix for each value of $p$, whereas with the assistance of expression \eqref{graphdiagonalexact} $E(\rho_p)$
is calculated via diagonalization of many $2^3\times2^3=8\times8$ dimensional matrices only.

%%%%%%%%%%%%%%%%%%%%%%%%%%%%%%%%%%%%%%%%%%%%%%%%%%%%%%%%%%%%%%%%%%%%%%%%%%%%%%%%%%%%%%%%%%%%%%%%%%%%%
\subsection{Independent thermal reservoirs on graph states}
\label{GAD}
In the case of Pauli maps the entanglement lower and upper bounds coincide, and deliver the exact entanglement.
However, this is not the case for general, non-Pauli, noise channels. The upper bound is given, as usual, by
convexity. The lower bounds must be optimized by appropriately 
choosing the LOCC operations. Here, we investigate and optimize
measurement strategies for
channel GAD, defined in Sec. \ref{Thermaldef}.

Observe that the Kraus operators defined in Eqs. (\ref{Kraus_AmpDamp}) satisfy the following: ${K_i}_0$ and ${K_i}_2$ commute with any $CZ$ operator, while for every $j\in\mathcal{V}$ different from $i$ it holds that $({K_i}_1\otimes \mathbf{1}_j).CZ_{ij}=CZ_{ij}.({K_i}_1\otimes Z_j)$ and
$({K_i}_3\otimes \mathbf{1}_j).CZ_{ij}=CZ_{ij}.({K_i}_3\otimes Z_j)$. Based on this, one can perform the factorization in equation
(\ref{commutation}) and apply this way the formalism described in Sec. \ref{sec:general}.

In what follows we focus on two main limits of channel GAD discussed in Sec. \ref{Thermaldef}:  the purely-dissipative limit $\overline{n}=0$ (amplitude damping), and the purely-difusive limit  $\overline{n}\rightarrow\infty$, $\gamma\rightarrow0$, and
$\overline{n}\gamma=\Gamma$.

\subsubsection{Graph states under zero-temperature dissipation}

\begin{figure}[!hb]
\centering
\includegraphics[scale=0.3]{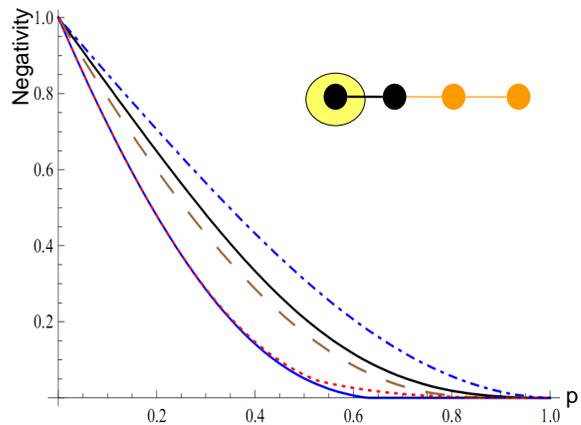}
\caption{(Color online) Negativity vs $p$ for a 4 qubit linear
cluster subjected to the amplitude damping channel and for the
partition displayed in the inset. Solid (upper) black curve: exact
entanglement; Solid (lower) blue curve: lowest lower bound LLB
(obtained by tracing out the flags); Dashed-dotted blue curve:
upper bound (obtained by convexity);
Dotted red curve: LB(0) (obtained by measuring the flags in the
$Z$ basis); Dashed brown curve: LB($\pi/4$) (obtained by measuring
the flags in the $X$ basis).} \label{4qubit_AmpDamp}
\end{figure}

We consider a four qubit linear
(1D) cluster state subjected
to the AD map and study the decay of entanglement in the
partition consisting of the first qubit versus the rest shown in the inset of Fig. \ref{4qubit_AmpDamp}.
Along with the exact calculation of entanglement via 
brute-force diagonalization of the partially-transposed matrices, the lowest lower bound LLB \eqref{lowernonoptimal}, obtained by tracing out the flags, and the
upper bound \eqref{uppernonoptimal}, obtained from convexity, are plotted. In addition, the tightness of the lower bounds \eqref{lower} obtained by the flag measurements can be scanned as a function of the measurement bases.

%%%%%%%%%%%%%%%%%%%%%%%%%%%%%%
\begin{figure}[t]
\centering
\includegraphics[scale=0.3]{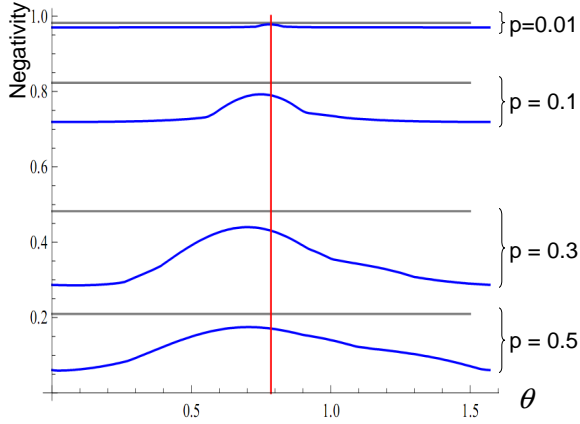}
\caption{(Color online) Lower bound LB($\theta$) to the negativity
as a function of the angle $\theta$ in the measurement basis, for
the same situation as in Fig. \ref{4qubit_AmpDamp}, and for fixed
values of $p$. Each value $p = 0.01, 0.1, 0.3, 0.5$ has two curves
associated to it. The horizontal (gray) straight line  represents
the exact entanglement at each $p$, while the blue (black) curve
represents the bound LB($\theta$) at this $p$. The red line
(vertical) corresponds to $\theta = \frac{\pi}{4}$, {\it i.e.}
measurements in the basis  $\{\ket{+},\ket{-}\}$.}
\label{4qubit_AmpDamp_Optimal_p=01}
\end{figure}

Based on observations about the behavior of the system under the AD
map we can guess good measurement strategies.
For example, examination of the initial state reveals that at $p=0$ each of
the non-boundary qubits is in one of the states of the  basis $\{\ket{+},\ket{-}\}$; whereas at $p=1$, in one of the states  of
$\{\ket{0},\ket{1}\}$. We call the lower bound obtained from \eqref{lower}
through measurements in the basis $\{\ket{+},\ket{-}\}$ LB($\pi/4$), and  the one obtained from \eqref{lower} through measurements in $\{\ket{0},\ket{1}\}$ LB(0). The latter bounds are the two additional curves plotted in Fig. \ref{4qubit_AmpDamp}. We observe that LB(0) provides only a
slight improvement over the LLB, whereas LB($\pi/4$) appears to give a
significant one. This raises the obvious question of how to optimize the choice of measurement basis at each instant $p$ in the evolution.

As an illustration we consider lower bounds LB($\theta$) obtained from \eqref{lower} through orthogonal measurements composed of projectors $\ket{\theta+}=\cos\theta \ket{0}+\sin\theta \ket{1}$ and $\ket{\theta-}=-\sin\theta\ket{0}+\cos\theta\ket{1}$, and look for the angle $\theta$ that gives us approximately the largest value of LB($\theta$). This is certainly not the most general measurement scenario one may consider, but it gives one a hint on how to increase the tightness of the bounds. 

%%%%%%%%%%%%%%%%%%%%%%%%%%%%%%%%%%%%%%%%%%%%%%%%%%%%%%%%%%%%%%%

\begin{figure}[t]
\centering
\includegraphics[scale=0.3]{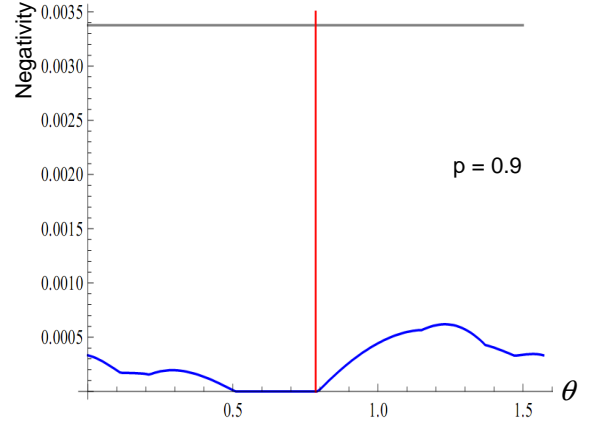}
\caption{(Color online) Same as Fig.
\ref{4qubit_AmpDamp_Optimal_p=01} for $p=0.9$.} \label{4qubit_AmpDamp_Optimal_p=09}
\end{figure}

Figs.
\ref{4qubit_AmpDamp_Optimal_p=01} and \ref{4qubit_AmpDamp_Optimal_p=09}
illustrate this idea. At fixed values of $p$, we have varied angle $\theta$ in the range  $[0,\pi /2]$. The entanglement given by LB($\theta$)
for each value of $\theta$ is compared with the exact
entanglement at the given $p$. In physical terms, we are taking
snapshots of the evolution of the system's entanglement at discrete time instants. 
The value of $\theta$ at each
instant $p$ that maximizes LB($\theta$) represents the
optimal measurement basis at that particular instant. As is clearly seen in Fig. \ref{4qubit_AmpDamp_Optimal_p=01},
%%%%%%%%%%%%%%%%  4 qubit GAD map, alpha=0.5,  %%%%%%%%%%%%%%%%%
\begin{figure}[b]
\centering
\includegraphics[scale=0.3]{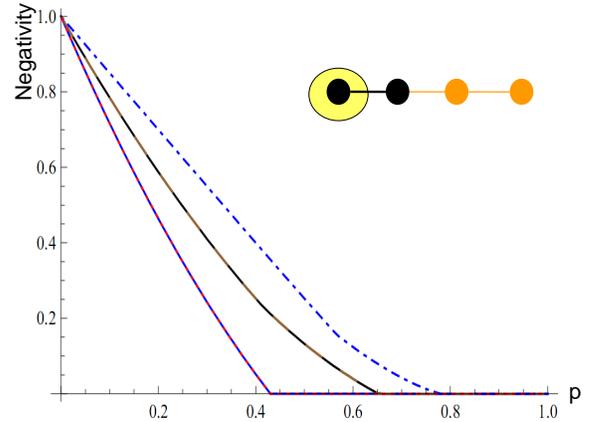}
\caption{(Color online) Negativity vs $p$ for a 4-qubit linear
cluster for the bipartition shown in the inset, subjected to the
generalized amplitude damping channel in the diffusive limit
$\overline{n}\rightarrow\infty$. The central curve corresponds to both the exact entanglement (solid black) and
LB($\pi/4$) (dashed brown), which coincide exactly. The lower curve corresponds to both LBB (solid blue) and LB(0)
(dotted red), which also coincide exactly. The upper curve is the upper bound
\eqref{uppernonoptimal} (dot-dashed blue).}
\label{4qubit_GenAmpDamp}
\end{figure}
 for small values of $p$ angles around
$\theta=\frac{\pi}{4}$ give the closest approximations to the exact
entanglement, in consistence with the significant improvement of LB($\pi/4$) over
the LLB observed in Fig. \ref{4qubit_AmpDamp}. For large values of $p$ though, the best approximations tend to be given by the angles away from $\theta=\frac{\pi}{4}$, as can be observed in Fig. \ref{4qubit_AmpDamp_Optimal_p=09}. It must still be kept in mind that none of these closest approximations
equals the exact entanglement of the state.
%%%%%%%%%%%%%%%%%%%%%%%%%%%%%%%%%%%%%%%%%%%%%%%%%%%%%%%%%%%%%%%
\subsubsection{Graph states under infinite-temperature difusion}

\begin{figure}[b]
\centering
\includegraphics[scale=0.32]{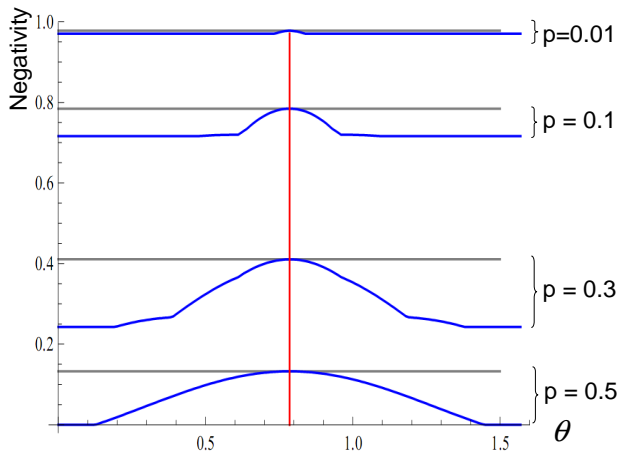}
\caption{(Color online) Lower bound LB($\theta$) to the negativity
as a function of the angle $\theta$ in the measurement basis, for
the same situation as in Fig. \ref{4qubit_GenAmpDamp}, and for
fixed values of $p$. Each value $p = 0.01, 0.1, 0.3, 0.5$ has two
curves associated to it. Again,  the horizontal (gray) straight
line represents the exact entanglement at each $p$, while the blue
(black) curve represents the bounds LB($\theta$) at the this $p$.
The red line (vertical) corresponds to $\theta = \frac{\pi}{4}$,
i.e. measurement in the basis $\{\ket{+},\ket{-}\}$.}
\label{4qubits_GenAmpDamp_Optimal}
\end{figure}

We now consider the purely-diffusive case of the GAD channel, where
each qubit is in contact with an independent bath of infinite
temperature. In Fig. \ref{4qubit_GenAmpDamp} we display the
entanglement evolution in a similar way as in Fig.
\ref{4qubit_AmpDamp}. Since in the purely-diffusive limit channel
GAD becomes a Pauli map, as was mentioned in the end of Sec.
\ref{Thermaldef}, bound LB($\pi/4$) yields the exact
entanglement.  LB(0) on the other hand coincides with the lowest
lower bound LLB.  The fact that in this case LB($\theta$) reaches
the exact entanglement at $\theta = \frac{\pi}{4}$ can also be seen
in a clearer way in Fig. \ref{4qubits_GenAmpDamp_Optimal}.
%%%%%%%%%%%%%%%%  4 qubit GAD map, alpha=0.5,  Optimal %%%%%%%%%%%%%%%%%

In Fig. \ref{4qubit_GenAmpDamp}  upper bound \eqref{uppernonoptimal} is plotted as well. Since in this case the channel is a Pauli channel, one would expect the upper bound to coincide with the exact entanglement as well. The fact that this does not occur is because, even though the noise itself is describable as a Pauli map, the plotted upper bound  has been calculated using the original Kraus decomposition of Eqs. \eqref{Kraus_AmpDamp}, which is not in a Pauli-map form.
For every given particular Kraus decomposition of a superoperator, the naive application of convexity always yields  UB through Eq. \eqref{uppernonoptimal}, but this needs not the tightest, for the Kraus decomposition of a superoperator is in general not unique. This observation  leads to a whole family of upper bounds for a given map. In the same spirit as with the lower bounds, one could in principle optimize the obtained UBs over all possible Kraus representations of the map.

%%%%%%%%%%%%%%%% Formalism %%%%%%%%%%%%%%%%%%%%%%%
\section{\label{sec:extlim}  Extentions and Limitations}

The framework developed here is not restricted to graph
states. The crucial ingredient in our formalism are to factor out all the entangling operations that act as local unitary transformations
with respect to the considered partition, and to redefine of the Kraus
operators acting on the state, reducing the entanglement evaluation problem to the boundary sub-system alone. Given an entangled state
and a prescription for its construction in terms of entangling
operations, useful bounds and exact expressions for the entanglement
can be readily obtained. As an example, a GHZ-like state $\left\vert
\psi\right\rangle = \alpha\left\vert
0\right\rangle ^{\otimes N}+\beta\left\vert 1\right\rangle ^{\otimes
N} $
can be operationally constructed by the sequential application of  maximally-entangling operation $CNOT_{ij}=\left\vert 0_{i}0_{j}\right\rangle
\left\langle 0_{i}0_{j}\right\vert +\left\vert
0_{i}1_{j}\right\rangle \left\langle 0_{i}1_{j}\right\vert
+\left\vert 1_{i}0_{j}\right\rangle \left\langle
1_{i}1_{j}\right\vert +\left\vert 1_{i}1_{j}\right\rangle
\left\langle 1_{i}0_{j}\right\vert $  to the product state $(\alpha\ket{0}+\beta\ket{1}) \otimes\left\vert 0\right\rangle \otimes...\otimes\left\vert
0\right\rangle $ such that $\left\vert \psi\right\rangle =
{\displaystyle\bigotimes\limits_{i=1}^{N-1}} CNOT_{i,i+1}(\alpha\ket{0}+\beta\ket{1}) \otimes\left\vert 0\right\rangle \otimes...\otimes\left\vert
0\right\rangle $.
Using our techniques and the permutation symmetry of the state it can be seen that, for GHZ-like states as above undergoing the previously discussed noise processes, the entanglement evaluation  in any
bipartition can be reduced to that of a two qubits system.
It is also important to mention that the techniques presented here can also be straightforwardly extended to higher-dimensional graph states \cite{highD}.

In addition, it is important to mention that all bounds developed so far
can in fact also be exploited to follow the entanglement evolution
when the system's initial state is a mixed graph-diagonal state.
This is simply due to the fact that any graph-diagonal state as
\eqref{graphdiagonal} can be thought of as a Pauli map
$\Lambda_{GD}$ acting on a pure graph state:
\begin{eqnarray}
\label{graphdiagonalmap} \nonumber \rho_{GD}=\sum_{\nu} P_{\nu}
\ket{{G_{(\mathcal{V},\mathcal{C})}}_{\nu}}\bra{{G_{(\mathcal{V},\mathcal{C})}}_{\nu}}=\\
\sum_{\nu} P_{\nu}Z^{\nu}
\ket{{G_{(\mathcal{V},\mathcal{C})}}_{0}}\bra{{G_{(\mathcal{V},\mathcal{C})}}_{0}}Z^{\nu}=\Lambda_{GD}(\ket{{G_{(\mathcal{V},\mathcal{C})}}_{0}}).
\end{eqnarray}
Thus, the entanglement at any time $t$ in a system initially in a
mixed graph-diagonal state $\rho_{GD}$, and evolving under some map
$\Lambda$, is equivalent to that of an initial pure graph state
$\ket{{G_{(\mathcal{V},\mathcal{C})}}_{0}}$ whose evolution is ruled
by the composed map $\Lambda\circ\Lambda_{GD}$, where $\Lambda_{GD}$
is defined in \eqref{graphdiagonalmap}. When $\Lambda$ is itself a
Pauli map, then $\Lambda\circ\Lambda_{GD}$ is also a Pauli map and
the expression \eqref{graphdiagonalexact} for the exact entanglement
can be applied. For the cases where  $\Lambda$ is not a Pauli map
but the relations \eqref{commutation} are satisfied by its Kraus
operators, the relations \eqref{commutation} will also be  satisfied
by the composed map $\Lambda\circ\Lambda_{GD}$, so that all other bounds derived here also hold.

Furthermore, as briefly mentioned before, any
arbitrary state can be depolarized by some separable map towards a graph-diagonal
state without changing the diagonal elements in the
considered graph basis \cite{Duer03Aschauer05}. The latter, since
the entanglement of almost all states cannot increase under separable maps \cite{Vlad}, implies that all the lower bounds presented here
also provide lower bounds to the decay of the entanglement that, though in general far from tight, apply to {\it almost any arbitrary initial state subject to
any decoherence process}.

The gain in computational effort provided by the machinery presented
here diminishes with the ratio between the number of
particles in the boundary subsystem and the total number of particles. For example, for multipartitions such that the boundary system is the total system itself, or for entanglement quantifiers that do not refer to any multi-partition at all, our method yields no gain. An example of the latter are the entanglement measures that treat all parties in a system indistinguishably, some of which, as was mentioned in the introduction, have been studied in Refs. \cite{Guehne,Wunderlich}. These methods naturally complement with ours to offer a rather general and versatile toolbox for the study of the open-system dynamics of graph-state entanglement.

%%%%%%%%%%%%%%%%%%%%%%%%%%%%%%%%%%%%%%%%%%%%%%%%%%%%%%%%%%%%%%%%

\section{\label{sec:conclusion}Conclusions}
We have studied in detail a general method for computing the
entanglement of graph and graph-diagonal  states undergoing decoherence, introduced in Ref. \cite{dan1}. This method allows to drastically reduce the effort to compute the entanglement evolution of graph states in several physical scenarios.  We have given an explicit formula for the construction of the effective noise involved in the calculation for Pauli maps and extended the formalism to the case of independent baths at arbitrary temperature. Also, we have elaborated the formalism to construct non-trivial lower and upper bounds to the entanglement decay where exact results cannot be obtained from the
formalism itself.

Finally we would like to add that the necessary requirements on the noise channels for the method to apply do not prevent us from obtaining general results for a wide variety of realistic decoherence processes. Furthermore, the conditions required on the entanglement measures are satisfied by most quantifiers.

\begin{acknowledgements}
DC acknowledges financial support from the National
Research Foundation and the Ministry of Education of Singapore. LA acknowledges the ``Juan de la Cierva" program for financial
support. RC and LD acknowledge support from the Brazilian agencies CNPq and FAPERJ, and from the National Institute of Science and Technology for Quantum Information. AA is supported by the European PERCENT ERC Starting Grant and Q-Essence project, the Spanish MEC FIS2007-60182 and Consolider-Ingenio QOIT projects, Generalitat de Catalunya and Caixa Manresa.

\end{acknowledgements}
%%%%%%%%%%%%%%%%%%%%%%%%%%%%%%%%%%%%%%%%%%%%%%%%%%%%%%%%%%%%%%%%

\end{document}